\documentclass[preprint2]{aastex}

\newcommand{\beq}{\begin{equation}}
\newcommand{\eeq}{\end{equation}}
\newcommand{\bea}{\begin{eqnarray}}
\newcommand{\ena}{\end{eqnarray}}

\begin{document}

\title{On a possible GRB-supernova time sequence}

\author{Remo Ruffini, Carlo Luciano 
Bianco, Federico 
Fraschetti, She-Sheng Xue}
\affil{ICRA - International Center for Relativistic Astrophysic\\ Physics 
Department, University of Rome ``La Sapienza''}
\affil{Piazzale Aldo Moro 5, I-00185 Rome, Italy.}
\and
\author{Pascal Chardonnet}
\affil{Universit\'e de Savoie - LAPTH LAPP - BP110 - 74941 Annecy-le-Vieux 
Cedex, France}
\altaffiltext{1}{ruffini@icra.it}

\begin{abstract}
The data from the Chandra satellite on the iron emission lines in the afterglow of GRB 991216 are used to give further support for the EMBH theory, which links the origin of the energy of GRBs to the extractable energy of electromagnetic black holes (EMBHs), leading to an interpretation of the GRB-supernova correlation. Following the relative space-time transformation (RSTT) paradigm and the interpretation of the burst structure (IBS) paradigm, we introduce a paradigm for the correlation between GRBs and supernovae. The following sequence of events is shown as kinematically possible and consistent with the available data: a) the GRB-progenitor star $P_1$ first collapses to an EMBH, b) the proper GRB (P-GRB) and the peak of the afterglow (E-APE) propagate in interstellar space until the impact on a supernova-progenitor star $P_2$ at a distance $\le 2.69\times 10^{17}$ cm, and they induce the supernova explosion, c) the accelerated baryonic matter (ABM) pulse, originating the afterglow, reaches the supernova remnants $18.5$ hours after the supernova explosion and gives rise to the iron emission lines. Some considerations on the dynamical implementation of the paradigm are presented. The concept of induced supernova explosion introduced here specifically for the GRB-supernova correlation may have more general application in relativistic astrophysics.
\end{abstract}


We have seen in the previous two letters how the fit of the data from the RXTE \citep{cs00} and Chandra \citep{p00} satellites on the afterglow of the GRB~991216 offers a tool to determine the only two free parameters of the EMBH theory. This theory links the energy source of GRBs to the electromagnetic mass energy of black holes \citep{cr71}. We have also seen how this theory has consequences for the interpretation of the structure of GRBs \citep[the IBS paradigm, see][]{lett2}. The same analysis also yields information about the density and overall distribution of the baryonic matter in the remnant left over from the gravitational collapse of the GRB-progenitor star to an EMBH, see \citet{lett1} and \citet{lett2}, and see also \citet{rcvx01g}. Similarly it allows one to probe the density distribution of the interstellar matter around the newly formed EMBH (see e.g. \citet{CNR,lett5}).

The aim of this letter is to point out that the data on the iron lines from the Chandra satellite on the GRB~991216 \citep{p00} and similar observations from other sources \citep{p99b,a00,p00} make it possible to extend this analysis to a larger distance scale, possibly all the way out to a few light years, and consequently probe the distribution of stars in the surroundings of the newly formed EMBH. These considerations lead to a new paradigm for the interpretation of the supernova-GRB correlation.

That indeed a correlation between the occurrence of GRBs and supernova events exists has been established by the works of \citet{b99,g98b,g98c,g00,k98,p98a,p99,r99,vp00}. Such an association has been assumed to indicate that GRBs are generated by supernovae explosions \citep[see e.g.][]{k98}.

We propose that, if relativistic effects are properly taken into account, then an alternative, kinematically viable, explanation can be given of the supernova-GRB association. We again use GRB~991216 as a prototypical case. The same theoretical considerations have been also applied to other cases, including GRB~980425 and SN~1998BW  \citep{lett4}.

We focus on the detailed kinematical description of this GRB-supernova time-sequence process and outline a possible dynamical scenario. We introduce a process by which a massive GRB-progenitor star $P_1$ of mass $M_1$ undergoes gravitational collapse to an EMBH. During this process a dyadosphere is formed and subsequently the P-GRB and the E-APE of \citet{lett2} are generated in sequence \citep[see also][]{lett1}. They propagate and impact, with their photon and neutrino components, on a second supernova-progenitor star $P_2$ of mass $M_2$. Assuming that both stars were generated approximately at the same time, we expect to have $M_2 < M_1$. For a wide range of parameters, such a collision will not affect the star $M_2$ \citep[][]{cr01}. Under some special conditions of the thermonuclear evolution of the supernova-progenitor star $P_2$, the collision can {\em induce a supernova explosion}.

We assume that the star $P_2$ is close to the line of sight of the EMBH. We will see in the following that this gives an upper limit to the distance $D_{P_2}=2.69\times 10^{17}$~cm of the supernova-progenitor star $P_2$ from the EMBH. The location of the star $P_2$ will then be constrained between the transparency point of the P-GRB, $9.692\times 10^{13}$ cm \citep[see][]{lett1}, and the above upper limit, and will be a function of the angle subtended by the line of sight and the star $P_2$, as seen from the EMBH. The energy-momentum deposited by the GRB in the collision with the star is in the range $10^{39}\sim 10^{45}$ ergs \citep[see e.g.][]{lett7}.

Especially relevant to our model are the following data from the Chandra satellite \citep[see][]{p00}:\\
1) At the arrival time of 37 hr after the initial burst there is evidence of
iron emission lines for GRB~991216.\\
2) The emission lines are present during the entire observation period of $10^4$ sec. The iron lines could also have been produced earlier, before Chandra was observing. Thus the times used in these calculations are not unique: they do serve to provide an example of the scenario.\\
3) The emission lines appear to have a peak at an energy of $3.49 \pm 0.06$ keV which, at a redshift $z=1.00 \pm 0.02$ corresponds to an hydrogen-like iron line at 6.97 keV at rest.  This source does not appear to have any significant motion departing from the cosmological flow. The iron lines have a width of 0.23 keV consistent with a radial velocity field of $0.1c$.

From the theoretical slope of the afterglow, presented in \citet{lett2}, we see that the flux of the afterglow observed by Chandra is in excellent agreement with the general afterglow slopes. Clearly the iron lines are only a small fraction of the observed flux.

We assume the laboratory frame as an inertial system of reference in which both stars $P_1$ and $P_2$ are at rest. A second  asymptotic inertial reference frame is assumed in which the detector is at rest. Therefore a detector arrival time $t_a^d$ is defined related to the laboratory time $t$ by
\begin{equation}
t_a^d=\left(1+z\right)t_a
=\left(1+z\right)\left(t-\frac{r\left(t\right)}{c}\right) ,
\label{taddef}
\end{equation}
where $z$ is the cosmological redshift, which in the case of GRB~991216 is equal to $z=1.0$, and $r\left(t\right)$ is the radius of the expanding pulse at laboratory time $t$.
 
To explain the above observations we propose that the expansion of the accelerated baryonic matter \citep[ABM pulse see][]{lett1} relativistically expanding away from newly formed EMBH reaches $P_2$ with a delay in arrival time of 18.5 hr \citep[the details of the computation are given in][]{bcfrx01}. The associated afterglow then illuminates the expanding supernova shell, producing the observed iron emission lines. The Chandra satellite observations then offer the first data on such an induced-supernova-explosion \citep{p00}.

On the basis of the explicit computations of the different eras presented in \citet{lett1}, we make three key points:\\
1) An arrival time of 37 hr in the detector frame corresponds to a radial distance from the EMBH traveled by the ABM pulse of $2.69\times10^{17}$ cm in the laboratory frame \citep[see][]{lett1}.\\
2) It is likely that a few stars are present within that radius as members of a cluster. It has 
become evident from observations of dense clusters of star-forming regions that a stellar average density of typically $ 10^2 \mbox{pc}^{-3} $ \citep{btk00} should be expected. There is also the distinct possibility for this case and other systems that the stars $P_1$ and  $P_2$ are members of a detached  binary system.\\
3) The possible observations at different wavelengths \citep{lett4} crucially depend on the relative intensities between the GRB and the supernova as well as on the value of the distance and the redshift of the source.

\begin{figure}
\epsscale{0.7}
\plotone{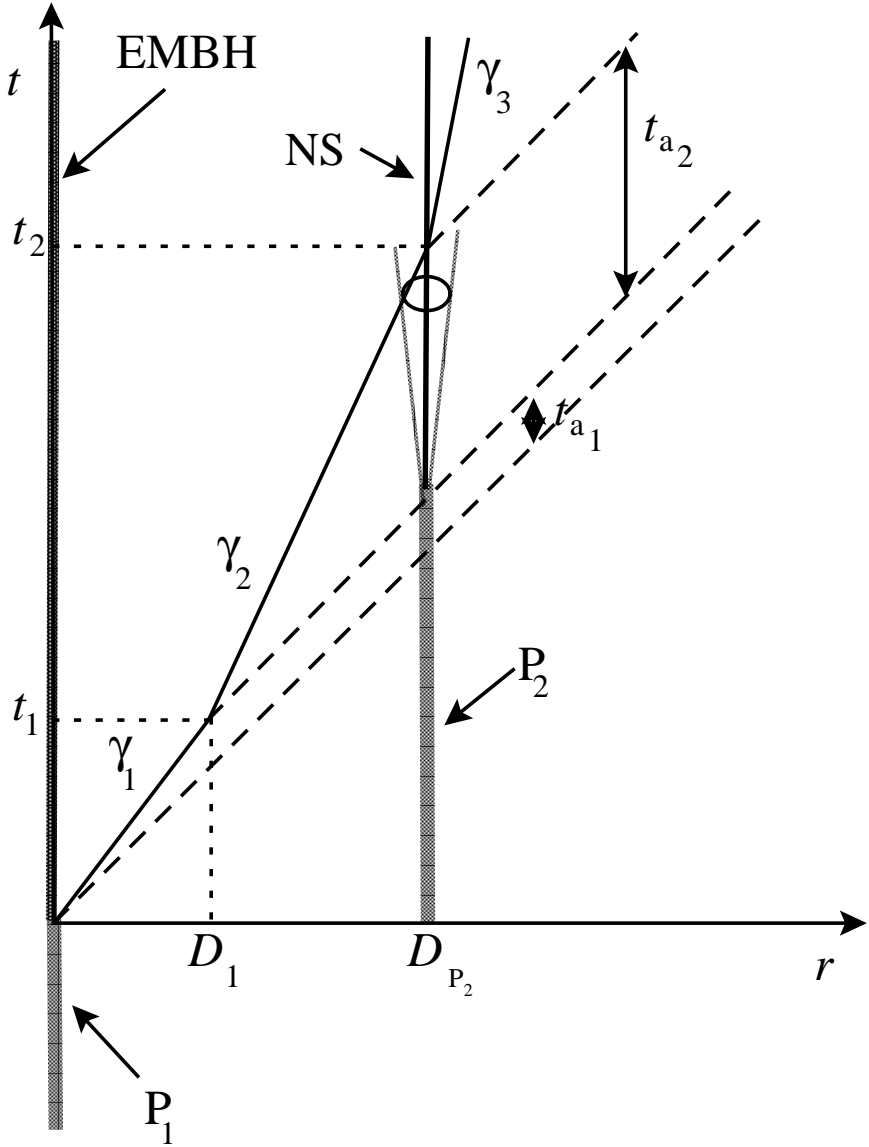}
\caption{A qualitative simplified spacetime diagram (in arbitrary units) illustrating the GSTS paradigm. The EMBH, originating from the gravitational collapse of a massive GRB-progenitor star $P_1$,  and the massive supernova-progenitor star $P_2$-neutron star ($P_2$-NS) system, separated by a radial distance $D_{P_2}$, are assumed to be at rest in in the laboratory frame. Their worldlines are represented by two parallel vertical lines. The supernova shell moving at $0.1c$ generated by the $P_2$-NS transition is represented by the dotted line cone. The solid line represents the motion of the pulse, as if it would move with an ``effective'' Lorentz factor $\gamma_1\simeq 110.0$ during the eras reaching the condition of transparency. Similarly, the ``effective'' Lorentz factor $\gamma_2\simeq 5.82$ applies during era IV up to the collision with the $P_2$-NS system. An ``effective'' Lorentz factor $\gamma_3 < 2$ occurs during era V after the collision as the nonrelativistic regime of expansion is reached (see \citet{lett6aa}). The dashed lines at 45 degrees represent signals propagating at speed of light.}
\label{Fig1}
\end{figure}

In order to reach an intuitive understanding of these complex computations we present a schematic very simplified diagram (not to scale) in Fig.~\ref{Fig1}.

We now describe the specific data of this GRB-supernova time-sequence (GSTS) paradigm:

1) The two stars $P_1$ and $P_2$ are separated by a distance $D_{P_2} = 2.69\times10^{17}$ cm in the laboratory frame, see Fig.~\ref{Fig1}. Both stars are at rest in the inertial laboratory frame. At laboratory time $t=0$ and at comoving time $\tau=0$, the gravitational collapse of the GRB-progenitor star $P_1$ occurs. The initial emission of gravitational radiation or a neutrino burst from the event then synchronizes the arrival times $t_a=0$ for the supernova-progenitor star $P_2$ and $t_a^d=0$ for the distant observer at rest with the detector. The electromagnetic radiation emitted by the gravitational collapse process is instead practically zero, by comparison, due to the optical thickness of the material at this stage \citep{brx00}.

2) From the determination of the parameters obtained in \citet{lett2} and the computations in \citet{rswx99,rswx00}, at laboratory time $ t_1 =3.295 \times 10^3 $ s and at a distance from the EMBH of $D_1=9.692\times 10^{13}$ cm, the condition of transparency for the pair-electromagnetic-baryon (PEMB) pulse is reached and the P-GRB is emitted, see Fig.~\ref{Fig1}, \citet{lett1} and \citet{lett2}. This time is recorded in arrival time at the detector $t_{a_1}^d = 0.1361 $ sec, and, at $P_2$, at $t_{a_1}=6.805\times 10^{-2}$ sec.

The fact that the PEMB pulse in an arrival time of 0.1361 sec covers a distance of $9.692\times10^{13}$ cm gives rise to an apparent superluminal effect. In order to clarify this apparent inconsistency we have introduced, from the computed values of $t_1$ and $t_{a_1}^d$, an ``effective'' Lorentz gamma factor 
by Eq.(\ref{gammaeff})
\begin{equation}
\gamma \equiv \sqrt{\frac{t}{2t_a^d}}\ ,
\label{gammaeff}
\end{equation}
which gives $\gamma_1\simeq 110.0$. Then we can straightforwardly explain the difference between the two times as $t_a^d=t/\left(2\gamma^2\right)$, see Fig.~\ref{Fig1}.

This introduction of an ``effective'' Lorentz gamma factor has no predictive power, it can only be introduced a posteriori, as an  heuristic tool in order to draw the qualitative diagram in Fig. 1. In practice the entire integration of the equations must be accomplished taking into account all changes of the time varying Lorentz gamma factors and the corresponding space and time variables throughout each era \citep[see for details][]{bcfrx01,lett6aa}.

3) At laboratory time $ t = 1.653 \times 10^6 $ s and at a distance from the EMBH of $4.863\times 10^{16}$ cm in the laboratory frame, the peak of the E-APE is reached which is recorded at the arrival time $t_a= 11.86$ sec at $P_2$  and $t_a^d = 23.72$ s at the detector. This also gives rise to an apparent superluminal effect which can be also explained following the same arguments given above in point 2). This event has not been represented in Fig.~\ref{Fig1} in order not to confuse the image.

4) At a distance $D_{P_2} = 2.69 \times 10^{17} $ cm, the two bursts described in the above points 2) and 3) collide with the supernova-progenitor star $P_2$ at arrival times $t_{a_1}=6.805\times 10^{-2}$ s and $ t_a  = 11.86 $ s respectively. They can then induce the supernova explosion of the massive star $P_2$.

5) The associated supernova shell expands with velocity $0.1c$.

\begin{figure}
\plotone{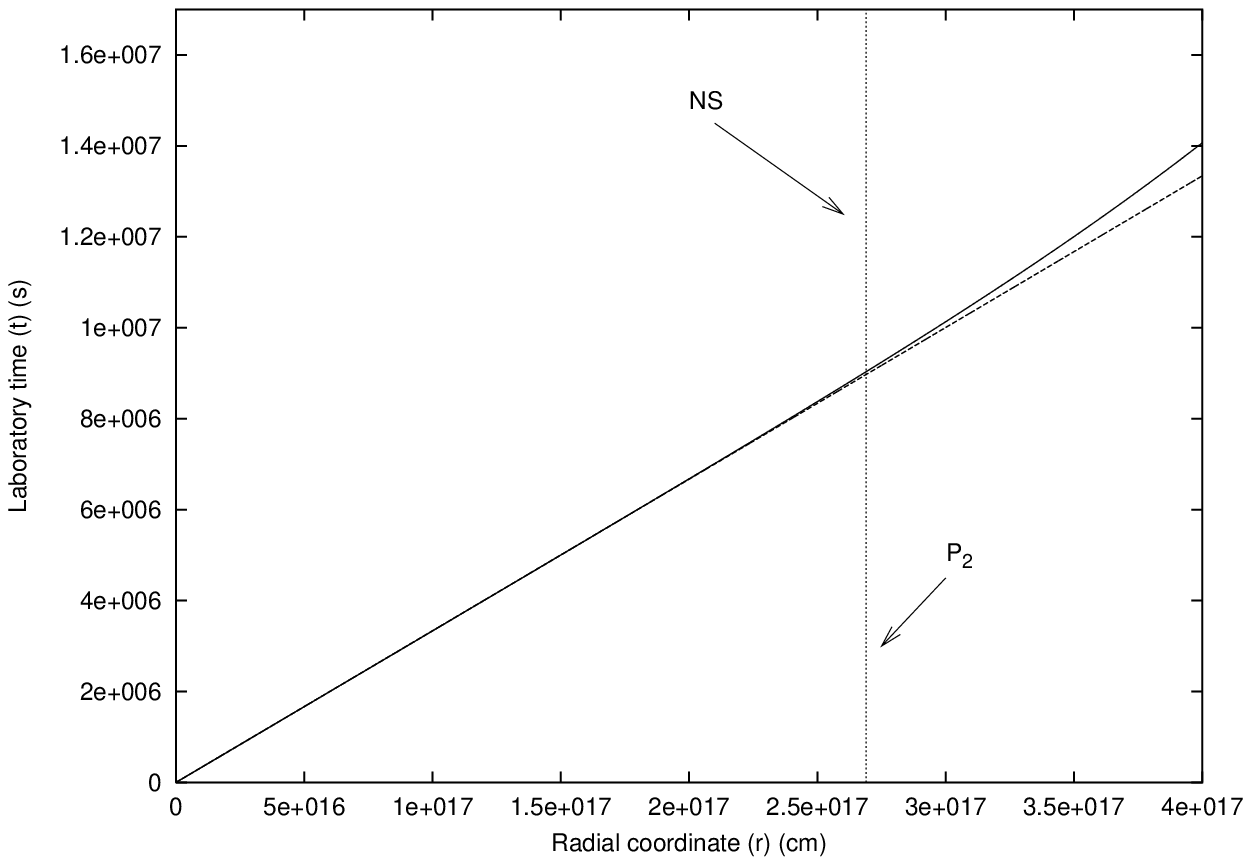}
\caption{The spatial radial coordinates of the P-GRB and the peak of the afterglow radiation flux (represented by a single dotted line, in this approximation), and of the pulse (represented by the solid line), as computed through the different eras presented in \citet{lett1}, using the results of \citet{rswx99,rswx00,bcfrx01}, are given as a function of the time in the laboratory frame. The vertical line corresponds to the radial position of the supernova-progenitor massive star $P_2$ $D_{P_2}=2.69\times 10^{17}$ cm, which undergoes supernova explosion after the collision with the P-GRB and the peak of the afterglow. The delay between the arrival time of the P-GRB, traveling at the speed of light, and the pulse, traveling with a Lorentz gamma factor $\gamma\simeq 3.38$ \citep[see][]{lett1} at the moment of collision with the supernova-progenitor star $P_2$, is $t_{a_2}=18.5$ hr, namely $t_{a_2}^d=37$ hr.}
\label{fig2}
\end{figure}

6) The expanding supernova shell is reached by the ABM pulse generating the afterglow  with a delay of $t_{a_2}=18.5$ hr in arrival time following the arrival of the P-GRB and the E-APE. This time delay coincides with the interval of laboratory time separating the two events, since the $P_2$ is at rest in the inertial laboratory frame (see Fig.~\ref{fig2}).

Again as explained above in point 2), this time delay can be interpreted a posteriori by introducing \citep{bcfrx01} an ``effective'' Lorentz gamma factor defined by
\begin{equation}
\gamma \equiv \sqrt{\frac{D_{P_2}}{2ct_{a_2}}}\ ,
\label{gammaeff2}
\end{equation}
which gives $\gamma \simeq 8.21$. Then we can heuristically visualize  the time delay as $t_{a_2}^d={D_{P_2}}/\left(2c\gamma^2\right)$. Clearly the results presented in Fig. 2 do follow the  complete integration of the equations of motion of the system through each different era defined in \citet{lett1}.

The ABM pulse will have travelled in the laboratory frame a distance $D_{P_2}-D_1\simeq D_{P_2}=2.69\times 10^{17}$ cm in a laboratory time $t_2-t_1 \simeq t_2=9.02\times 10^6$ s (neglecting the supernova expansion).

This again  gives rise to an apparent superluminal effect which can be interpreted heuristically, as in point 2), as a relativistic motion of the ABM pulse with an ``effective'' Lorentz factor of $\gamma_2\simeq 5.82$, see Fig. 1. The era IV extends from the point of transparency \citep[point 4 in Fig.~1 of][]{lett1} all the way to the collision of the pulse with the supernova shell, which occurs at $\gamma\simeq 3.38$. By this time the supernova shell has reached a dimension of $1.997\times 10^{14}$ cm, which is consistent with the observations from the Chandra satellite.

In the above considerations of GRB~991216 the supernova remnant has been assumed to be close to, but not exactly along, the line of sight extending from the EMBH to the distant observer. However, such a case should exist for other GRBs and would lead to an observation of iron absorption lines as well as to an increase in the radiation observed in the afterglow corresponding to the crossing of the supernova shell by the ABM pulse. In fact, as the ABM pulse engulfs the baryonic matter of the remnant, above and beyond the normal interstellar medium baryonic matter, the conservation of energy and momentum implies that a larger amount of internal energy is available and radiated in the process \citep{lett5}. This increased energy-momentum loss will generally affect the slope of the afterglow decay, approaching more rapidly a nonrelativistic expansion phase \citep[details are give in][]{lett6aa}.

If we now turn to the possibility of dynamically implementing the scenario, there are, at least, three different possibilities:\\
1) Particularly attractive is the possibility that a massive star $P_2$ has rapidly evolved during its thermonuclear evolution to a white dwarf \citep[see e.g.][]{c78}. It it then sufficient that the P-GRB and the E-APE implode the star sufficiently as to reach a central density above the critical density for the ignition of thermonuclear burning. Consequently, the explosion of the star $P_2$ occurs, and a significant fraction of a solar mass of iron is generated. These configurations are currently generally considered precursors of some type I supernovae \citep[see e.g.][ and references therein]{f97}.\\
2) Alternatively, the massive star $P_2$ can have evolved to the condition of being close to the point of gravitational collapse, having developed the formation of an iron-silicon core, type II supernovae. The above transfer of energy momentum from the P-GRB and the E-APE may enhance the capture of the electrons on the iron nuclei and consequentely decrease the Fermi energy of the core, leading to the onset of gravitational instability \citep[see e.g.][ p. 270 and followings]{b91}. Since the time for the final evolution of a massive star with an iron-silicon core is short, this event will require a perhaps unlikely coincidence.\\
3) The pressure wave may trigger massive and instantaneous nuclear burning process, with corresponding changes in the chemical composition of the star, leading to the collapse.

The GSTS paradigm has been applied to the case of the correlation between SN 1998bw and GRB 980425, which, with a redshift of 0.0083, is one of the closest and weakest GRBs observed. In this case the EMBH appears to have a significantly lower value of the parameter $\xi$ but the validity of the GSTS paradigm presented here is fully confirmed \citep[see][]{lett4}.

The GSTS paradigm and the concept of {\em induced supernova explosion}, which we have introduced for the collapse to an EMBH, may play a role also in the case of a collapse of a white dwarf core to a neutron star in a binary system. It may solve the long lasting problem of the almost equality of neutron star masses observed in some binary pulsars \citep[see e.g.][]{tw89}.

\acknowledgments

We thank three anonymous referees and J.~Wilson for their remarks, which have improved the presentation of this letter

\clearpage

\onecolumn

\end{document}